\documentstyle[12pt]{article}
\newcommand{\beq}{\begin{equation}}
\newcommand{\eeq}{\end{equation}}
\newcommand{\beqa}{\begin{eqnarray}}
\newcommand{\eeqa}{\end{eqnarray}}
\newcommand{\ba}{\begin{array}}
\newcommand{\ea}{\end{array}}
\newcommand{\CR}{\nonumber \\}
\newcommand{\pa}{\partial}
\newcommand{\A}{\alpha}
\newcommand{\B}{\beta}

\newcommand{\La}{\Lambda}

\newcommand{\lm}{\lambda}

\newcommand{\rep}{{\cal R}}
\newcommand{\pre}{{\cal F}}
\newcommand{\half}{{1\over 2}}
\newcommand{\Ba}{\bar{a}}
\newcommand{\Bb}{\bar{b}}
\newcommand{\cF}{{\cal F}}
\newcommand{\cR}{{\cal R}}

\newcommand{\LM}{\Lambda}
\newcommand{\ep}{\epsilon}

\begin{document}

\begin{titlepage}
\null
\begin{flushright}
hep-th/9712018   \\
UTHEP-375  \\
December, 1997  \\
\end{flushright}
\vspace{0.5cm}
\begin{center}
{\Large \bf
A-D-E Singularity and Prepotentials \\ 
in N=2  Supersymmetric Yang-Mills Theory
\par}
\lineskip .75em
\vskip2.5cm
\normalsize
{\large Katsushi Ito and Sung-Kil Yang}
\vskip 1.5em
{\large \it Institute of Physics, University of Tsukuba, Ibaraki 305, Japan}
\vskip3cm
{\bf Abstract}
\end{center} \par
We calculate the instanton corrections in the effective prepotential
for $N=2$ supersymmetric Yang-Mills theory with all A-D-E gauge groups from 
the Seiberg-Witten geometry constructed out of the spectral curves
of the periodic Toda lattice. The one-instanton contribution is determined 
explicitly by solving the Gauss-Manin
system associated with the A-D-E singularity. Our results are in complete 
agreement with the ones obtained from the microscopic instanton calculations.

\end{titlepage}

\baselineskip=0.7cm

\section{Introduction}

According to Seiberg-Witten (SW) the exact solution describing the
Coulomb branch of $N=2$ supersymmetric gauge theory in four dimensions is
formulated in terms of a Riemann surface together with a specific
differential $\lambda_{SW}$ \cite{SW}. 
A non-trivial test of the SW type solution is to 
check if the instanton corrections to the prepotential obtained from the 
non-perturbative
geometric description agree with the microscopic instanton calculations.
So far the tests have been carried out successfully for $N=2$ Yang-Mills
theory as well as $N=2$ QCD with various classical gauge groups 
\cite{instanton}-\cite{KlNoPh}.
For the classical groups Riemann surfaces are described in terms of  
hyper-elliptic curves. Toward the generalization to the case of the
exceptional gauge groups, several groups attempted the hyper-elliptic type
description by embedding the exceptional groups to certain classical gauge 
groups \cite{hyperell}.

Martinec and Warner \cite{MW} proposed another type of Riemann surfaces 
based on the spectral curve of the periodic Toda lattice for any gauge groups. 
In the case of  the exceptional gauge groups, however, the spectral curves 
are different from the hyper-elliptic ones and have different 
strong-coupling physics. Recent analysis of instanton expansions
in the case of exceptional gauge
groups $G_2$ \cite{It} and $E_6$ \cite{Ghe}
suggests that only the spectral curves yield the result consistent with the 
microscopic instanton calculus.

Our purpose in this paper is to perform the weak-coupling analysis of SW
geometry, and 
in particular to examine the one-instanton correction to the prepotential 
for $N=2$ Yang-Mills theory with 
$E_6,\, E_7$ and $E_8$ gauge groups based on the spectral curves.
In fact our
procedure applies systematically to all A-D-E gauge groups. As will be seen,
this is by virtue of the fact that the Picard-Fuchs equations for the $N=2$
SW period integrals are equivalent to the Gauss-Manin system for 
two-dimensional A-D-E topological Landau-Ginzburg (LG) models and the scaling 
relation for $\lambda_{SW}$, which is reported in our previous communication
\cite{IY}.

In section 2 the Picard-Fuchs equations formulated  in terms of 
the flat coordinates are presented. We derive the solutions of these 
Picard-Fuchs equations in the weak-coupling region in section 3. 
We then analyze in detail the one-instanton contribution and compare with 
the microscopic calculations in section 4. Our results are in complete 
agreement with the ones obtained from the microscopic instanton calculations.
As a by-product it is found that the Gauss-Manin system for the A-D-E
singularity has logarithmic solutions. This is described in section 5. 
Finally section 6 is devoted to discussions and conclusions.

\section{Picard-Fuchs equations}

We start with briefly summarizing a general scheme for SW curves. Martinec and
Warner proposed that the SW Riemann surface for $N=2$ Yang-Mills theory with
a simple group $G$ is given by a spectral curve of the periodic Toda lattice
associated with the dual group $G^\vee$ \cite{MW}. 
For $G=ADE$, in particular,
$G=G^\vee$ and, given any irreducible representation ${\cal R}$ of $G$, the
spectral curve is written in terms of the characteristic polynomial for 
${\cal R}$
\beq
P_G^{\cal R} \left( x, u_1,\cdots , u_{r-1},u_r+z+\frac{\mu^2}{z} \right) =0,
\label{curve}
\eeq
which is of degree ${\rm dim}\, {\cal R}$ in $x$, and $z$ is a spectral 
parameter \cite{LW}. Here each Casimir $u_i$ 
$(i=1,\cdots ,r;\, r={\rm rank}\, G)$ has degree $q_i=e_i+1$  where $e_i$
is the $i$-th exponent of $G$.
Throughout this paper, we denote the quadratic Casimir by $u_1$ and the
top Casimir of degree $h$ by $u_r$ with $h$ being the dual Coxeter number 
of $G$; $h=r+1, 2r-2, 12, 18,30$ for $G=A_r, D_r, E_6, E_7, E_8$, respectively.
When we identify (\ref{curve}) with the $N=2$ SW curve the Casimirs 
are regarded
as gauge invariant moduli parameters in the Coulomb branch and 
$\mu^2=\La^{2 h}/4$ with the dynamical scale $\La$. The meromorphic SW
differential is given by
\beq
\lambda_{SW}={x \over 2\pi i} \, {dz \over z}.
\label{swd}
\eeq

To describe the moduli space of the Coulomb branch we adopt the
flat coordinate system $(t_1, t_2, \cdots ,t_r)$ developed 
in the A-D-E singularity theory instead of the conventional Casimir
coordinates $(u_1, u_2, \cdots ,u_r)$. The
coordinate transformation is read off from the residue integral
\beq
t_i=c_i \oint dx W_G^\rep(x,u)^{e_i \over h}, \hskip10mm i=1,\cdots, r
\label{flat}
\eeq
with a suitable constant $c_i$ \cite{EY},\cite{EYY}. Here $W_G^\rep$ is 
obtained by solving (\ref{curve}) with respect to $u_r$
\beq
z+{\mu^2 \over z}+u_r=\widetilde W_G^\rep (x,u_1,\cdots ,u_{r-1}),
\label{solve}
\eeq
and setting
\beq
W_G^\rep (x,u_1,\cdots, u_r)=\widetilde W_G^\rep (x,u_1,\cdots, u_{r-1})-u_r.
\label{sp}
\eeq
Notice that the overall degree of $W_G^\rep$ is equal to $h$.
The flat coordinates $t_i$ are expressed as polynomials in $u_i$.
By degree counting we see 
\beq
{\partial \over \partial t_r} =
\sum_{i=1}^r {\pa u_i \over \pa t_r}{\partial \over \partial u_i}
=-{\partial \over \partial u_r},
\eeq
 and hence
\beq
{\partial W_G^\rep \over \partial t_r}=1.
\label{identity}
\eeq

Introducing the flat coordinates for the $N=2$ moduli space
is motivated by the fact that 
$W_{A_r}^{\bf r+1}$ and $W_{D_r}^{\bf 2r}$ are the well-known LG
superpotentials for the $A_r$- and $D_r$-type topological minimal models
respectively \cite{DVV}. 
Moreover it is demonstrated explicitly that $W_{E_6}^{\bf 27}$
describes precisely the $E_6$ topological LG model \cite{EY}.
In the following, therefore, we are led 
to assume that $W_{E_7}^{\bf 56}$ and $W_{E_8}^{\bf 248}$
give the single-variable version of the LG superpotentials for the
$E_7$ and $E_8$ models respectively.

In terms of the flat coordinates, the
Picard-Fuchs equations for the SW period integrals $\Pi =\oint \lambda_{SW}$
take a succinct form \cite{IY}
\beqa
&& {\cal L}_0 \Pi \equiv
\left( \sum_{i=1}^{r}q_{i}t_{i}{\pa\over \pa t_{i}}-1 \right)^2\Pi
-4\mu^2 h^{2}{\pa^{2}\Pi\over \pa t_{r}^2}=0, \CR
&& {\cal L}_{ij} \Pi \equiv
{\pa^{2}\Pi\over \pa t_{i}\pa t_{j}}
-\sum_{k=1}^r {C_{i j}}^{k}(t) {\pa^{2}\Pi \over \pa t_{k}\pa t_{r}}=0,
\label{PFflat}
\eeqa
where ${C_{i j}}^{k}(t)$ are the three-point functions in the 
two-dimensional
A-D-E topological LG models. We refer to the first equation
of (\ref{PFflat}) as the scaling equation and the second ones as the
Gauss-Manin system. 
Exactly the same form of the Gauss-Manin system has originally appeared 
in the study of the A-D-E singularity \cite{KS},\cite{No}, and become 
relevant in the context of two-dimensional topological gravity \cite{EYY}.

In \cite{MW} it is argued that the spectral curve (\ref{curve}) describes the
same physics of the Coulomb branch of $N=2$ A-D-E Yang-Mills theory 
{\it irrespective} of the representation $\rep$. 
Thus we expect that the explicit
form of the Picard-Fuchs equations (\ref{PFflat}) does not depend on $\rep$.
This also implies that, independently of $\rep$, the superpotential $W_G^\rep$ 
gives rise to the same topological field theory results with the standard 
A-D-E topological LG models. As an instructive example let us quote the
result for the representations ${\bf 5}$ and ${\bf 10}$ of $A_4$ \cite{EY}.
The characteristic polynomials for ${\bf 5}$ and ${\bf 10}$ are 
\beqa
P_{A_4}^{\bf 5}&=& x^5-u_1 x^3-u_2 x^2-u_3 x-u_4,  \CR
P_{A_4}^{\bf 10}
&=& {1\over 4}(p_1^2p_2-q_1^2)+u_4q_1-u_4^2  \CR
&=& x^{10}-3u_1x^8-u_2x^7+3(u_1^2+u_3)x^6+(2u_1u_2+11u_4)x^5
+(-2u_1u_3-u_1^3-u_2^2)x^4 \CR
&& +(-4u_1u_4-4u_2u_3-u_1^2u_2)x^3+(-u_1^2u_3-4u_3^2+u_1u_2^2+7u_2u_4)x^2 \CR
&& +(4u_3u_4+u_2^3+u_1^2u_4)x-u_4^2+u_2^2u_3-u_1u_2u_4,
\eeqa
where
\beqa
&&q_1=11 x^5-4 u_1 x^3+7 u_2 x^2+(u_1^2+4 u_3) x-u_1 u_2, \CR
&&p_1=5 x^3-u_1 x+u_2,\CR
&&p_2=5 x^4-2 u_1 x^2+4 u_2 x+u_1^2+4 u_3.
\eeqa
The superpotentials are obtained as
\beqa
&& W_{A_4}^{\bf 5}=P_{A_4}^{\bf 5},  \CR
&& W_{A_4}^{\bf 10}=\half (q_1\pm p_1\sqrt{p_2})-u_4.
\eeqa
Applying (\ref{flat}) we observe that both superpotentials yield the
identical formula for the flat coordinates
\beq
 t_1=-\omega^3 u_1, \hskip5mm
 t_2=-\omega^2 u_2, \hskip5mm
 t_3=-\omega \left( u_3+{u_1^2\over 5}\right), \hskip5mm
 t_4=-\left( u_4+{u_1u_2\over 5}\right)
\eeq
with $\omega =1/5^{1/5}$. Furthermore it is shown that the chiral ring
obtained from $W_{A_4}^{\bf 10}$ has the same structure with that from
$W_{A_4}^{\bf 5}$.

These $\rep$-independent results may be
considered as the materialization of the universality of the special Prym 
variety, which is a complex rank $G$-dimensional sub-variety of the
Jacobian of the curve, known in the theory of spectral curves \cite{spect}.


Finally we note that when dealing with (\ref{PFflat}) one may need to know 
the explicit form of ${C_{i j}}^{k}(t)$.
It is well-known that there exists the free energy
$F(t)$ such that $C_{ijk}(t)= \pa^3 F(t)/ \pa t_i \pa t_j \pa t_k$ where
$C_{ijk}(t)={C_{i j}}^{\ell}(t)\eta_{\ell k}$ with 
$\eta_{\ell k} =\delta_{e_\ell +e_k, h}$,  and 
$C_{ijr}(t)=\eta_{ij}$ \cite{DVV}. For the $A_4$ model the free
energy is given by
\beq
F_{A_4}(t)=t_2t_3t_4+{t_1t_4^2 \over 2}+{t_3^3 \over 6}-{t_2^4 \over 12}
-{t_1t_2^2t_3 \over 2}-{ t_1^2t_3^2\over 4}+{t_1^3t_2^2 \over 6}
-{t_1^6 \over 120}.
\eeq
When we analyze the instanton expansion, however, we can proceed without
using ${C_{i j}}^{k}(t)$ explicitly as will be seen in the next section.

\section{Solutions of the Picard-Fuchs equations}

In the semi-classical weak-coupling region where $\mu^2 \ll 1$,
we can find solutions of (\ref{PFflat}) explicitly. Let $e^{(b)}$ 
$(b=1,\cdots ,{\rm dim}\, \rep)$ be a root of the characteristic polynomial,
then
\beq
P_G^{\cal R}(x,u_1,\cdots ,u_r)=\prod_{b=1}^{{\rm dim}\, \rep}(x-e^{(b)})
\label{charapol}
\eeq
and 
\beq
e^{(b)}=(\lambda^{(b)}, \bar a),
\label{eq:15}
\eeq
where $\lambda^{(b)}$ are the weights of $\rep$, $(\; ,\; )$ stands for the 
inner product and
\beq
\bar a =\sum_{i=1}^r \bar a_i \alpha_i
\label{vev}
\eeq
with $\alpha_i$ being the simple roots of $G$.

For our discussions it suffices to take generic values of $\bar a_i$ at which
$e^{(b)} \not= e^{(b')}$ for $b\not= b'$. It follows that $e^{(b)}$ also 
satisfy
\beq
W_G^{\cal R}(e^{(b)}, u_1,\cdots ,u_r)=0, \hskip10mm
\pa_x W_G^{\cal R}(e^{(b)}, u_1,\cdots ,u_r)\not= 0,
\eeq
where the second equation is easily derived from 
\beq
\pa_x P_G^{\cal R}(e^{(b)}, u_1,\cdots ,u_r)\not= 0
\eeq
with the aid of the theorem on implicit function.

For simplicity let us denote $W_G^{\cal R}$ as $W$ henceforth. We 
obtain from (\ref{swd}) and 
(\ref{sp}) that
\beq
\lambda_{SW}={1\over 2\pi i} {x W' \over \sqrt{W^2-4\mu^2}}\, dx,
\eeq
where $'=\pa/\pa x$. In the classical limit $\mu^2 \rightarrow 0$,
$\lambda_{SW}$ has a simple pole at $x=e^{(b)}$. When we turn on 
$\mu^2 \not= 0$ the pole splits into two branch points at $x=e^{(b)}_{\pm}$
where $e^{(b)}_+ -e^{(b)}_- = O(\mu)$. Fix a branch cut connecting
$e^{(b)}_+$ and $e^{(b)}_-$, and take a closed contour $C^{(b)}$
enclosing this branch cut anti-clockwise, then, expanding $\lambda_{SW}$
around $\mu^2 =0$, we may evaluate the period as
\beqa
\pi^{(b)}(t) &=& \oint_{C^{(b)}} \lambda_{SW} \CR
&=& \sum_{n\geq 0} {(\mu^2)^n \over (n!)^2} 
\left( {\partial \over \partial t_r} \right)^{2n} 
\oint_{e^{(b)}} {dx \over 2\pi i}{x W' \over W} \CR
&=& \sum_{n\geq 0} {(\mu^2)^n \over (n!)^2} 
\left( {\partial \over \partial t_r} \right)^{2n} e^{(b)}(t),
\label{Cperiod}
\eeqa
where we have used (\ref{identity}).

Our task now is to prove that $\pi^{(b)}(t)$ obtained above satisfies 
the Picard-Fuchs equations (\ref{PFflat}). We first point out that 
${C_{ij}}^k(t)$ are independent of $t_r$ because
\beq
{\partial \over \partial t_r} C_{ijk}
={\partial \over \partial t_i} 
\left( {\partial F(t) \over \partial t_j \partial t_k \partial t_r}\right)
={\partial \eta_{jk}\over \partial t_i}=0,
\eeq
which ensures 
\beq
[{\cal L}_{ij},\, \pa /\pa t_r]=0.
\label{eq:comm}
\eeq
Therefore, if the roots $e^{(b)}$ satisfy the Gauss-Manin system of
(\ref{PFflat}) so do $\pi^{(b)}$'s. 
In fact it is not difficult to show that $e^{(b)}(t)$ is a solution 
of the Gauss-Manin system. 
The proof goes as follows.
The derivative of $W(e^{(b)}(t),t)=0$ with respect to $t_i$ yields 
\beq
{\pa e^{(b)} \over \pa t_i}=-{\phi_i \over W'},
\label{1st}
\eeq
where the RHS is evaluated at $x=e^{(b)}(t)$ and
\beq
\phi_i(x,t)={\pa W(x,t) \over \pa t_i}, \hskip10mm i=1, \cdots ,r
\eeq
are chiral primary fields in the LG model. Notice that $\phi_r=1$ as is seen
{}from (\ref{identity}). The fields $\phi_i$ obey the algebra
\beq
\phi_i \phi_j =\sum_k {C_{ij}}^k(t) \phi_k +Q_{ij}W',
\label{2nd}
\eeq
where we have
\beq
Q_{ij}'={\pa^2 W \over \pa t_i \pa t_j}
\label{qij}
\eeq
by virtue of the flat coordinates. Differentiate again (\ref{1st}) with 
respect to $t_j$ to obtain
\beqa
{\pa^2 e^{(b)} \over \pa t_i \pa t_j}
&=& {(\phi_i\phi_j)' \over {W'}^2}-{Q_{ij}'\over W'}
-{W'' \over {W'}^3}\, \phi_i \phi_j  \CR
&=& \sum_k {C_{ij}}^k(t) 
\left( {\phi_k' \over {W'}^2}-{W'' \over {W'}^3} \phi_k \right),
\label{3rd}
\eeqa
where (\ref{1st}), (\ref{2nd}) and (\ref{qij}) 
have been utilized. Notice that
${C_{r\ell}}^k=C_{r\ell j}\eta^{jk}=\eta_{\ell j}\eta^{jk}={\delta_\ell}^k$,
and hence putting $i=\ell,\, j=r$ in (\ref{3rd}) gives
\beq
{\pa^2 e^{(b)} \over \pa t_\ell \pa t_r}
= {\phi_\ell' \over {W'}^2}-{W'' \over {W'}^3} \phi_\ell .
\eeq
Thus we have shown that the zeroes of the characteristic polynomial for
any irreducible representation of the A-D-E groups satisfy the Gauss-Manin
system (\ref{PFflat}) for A-D-E singularities.

For the fundamental representations of $A_r$ and $D_r$
this fact has been known to mathematicians since the forms of 
$W_{A_r}^{\bf r+1}$ and $W_{D_r}^{\bf 2r}$
are standard in the theory of isolated singularities of types $A_r$ and $D_r$
\cite{Oda}. \footnote[2]{We thank Kyoji Saito and Y. Yamada for informing 
this to us. Our proof here is based on Yamada's unpublished note (March, 1993)
on the $A_r$-type singularity.} Our present result not only extends the
foregoing observation to the case of $E$-type singularities, but finds out 
the intimate relationship of the polynomial invariants with the Gauss-Manin
system.

We next show that $\pi^{(b)}$ satisfy the scaling equation of (\ref{PFflat}).
For this purpose we recall the quasihomogeneous property of the superpotential
\beq
x W'+\sum_{i=1}^r q_it_i{\partial W \over \partial t_i} =hW.
\eeq
Putting $x=e^{(b)}$ here and using (\ref{1st}) it is immediate to see that
\beq
{\cal D}e^{(b)} \equiv 
\left( \sum_{i=1}^{r} q_it_i{\pa \over \pa t_i}-1 \right) e^{(b)}=0.
\eeq
{}From this and 
\beq
\left[ {\cal D}^2,\; {\pa^m \over \pa t_r^m} \right]
={\pa^m \over \pa t_r^m} (m^2h^2-2mh {\cal D}),
\eeq
one can explicitly check that (\ref{Cperiod}) 
indeed satisfy the scaling equation.

Finally we note that $\bar a_i(t)$ introduced in (\ref{vev}) determine
the classical values of the adjoint Higgs field $\Phi$ in the $N=2$ vector
multiplet
\beq
\Phi=\sum_{i=1}^r \bar a_i H_{i},
\eeq
where $H_i$ are the Cartan generators normalized as ${\rm Tr}\; H_iH_j=K_{ij}$
with $K_{ij}$ being the Cartan matrix. This means that the $A$-cycles $A_i$
on the Riemann surface (\ref{curve}) may be fixed by the relation
\beq
C^{(b)}=\sum_{i=1}^r (\lambda^{(b)}, \alpha_i)A_i.
\eeq
The SW periods along the $A$-cycles are thus given by
\beq
a_i(t) = \oint_{A_i} \lambda_{SW} , \hskip10mm i=1,\cdots ,r
\eeq
which have the weak-coupling expansion
\beq
a_i(t)=\bar a_i(t)+\sum_{k\ge 1} \tilde a_i^{(k)}(t) \La^{bk},
\label{aweak}
\eeq
where $b=2h$ is the coefficient of the one-loop beta function and
\beq
\tilde a_i^{(k)}(t)={1 \over 4^k (k!)^2}
\left( {\partial \over \partial t_r} \right)^{2k} \bar a_i(t).
\label{kthinst}
\eeq

\section{One-instanton corrections}

Let us now turn to the analysis of instanton expansions in A-D-E Yang-Mills
theory. 
It is known that the $N=2$ prepotential in the weak-coupling region of
the Coulomb branch takes the form
\beq
\pre (a)={\tau_{0}\over2} \sum_{i,j=1}^{r}a_{i}K_{ij} a_{j}+
{i\over 4\pi} \sum_{\alpha \in \Delta^+} (\alpha ,a)^2 
\ln {(\alpha ,a)^2 \over \La^2}+\sum_{k\ge 1} \pre_k(a) \La^{bk},
\label{prepo}
\eeq
where $a(t)=\sum_{i=1}^r a_i(t)\alpha_i$, 
$\tau_{0}$ is the bare coupling constant, 
$\pre_k(a)$ stands for the 
$k$-instanton contribution and $\Delta^+$ is a set of positive roots of $G$.

In order to evaluate $\pre_1(a)$ we invoke the scaling relation \cite{MIEY}
\beq
\sum_{i=1}^r a_i {\pa \pre (a) \over \pa a_i}-2\pre (a)
={ib \over 2\pi}u_1,
\label{scale}
\eeq
where the quadratic Casimir $u_1$ and $\bar a_i$ are related through
\beq
u_1={1\over 2h} \sum_{\alpha \in \Delta^+} (\alpha , \bar a)^2
={1 \over 2}{\rm Tr}\, \Phi^2.
\label{2ji}
\eeq
We substitute the weak-coupling expansion (\ref{prepo}) together with 
(\ref{aweak}) into (\ref{scale}). Making use of (\ref{2ji}) and
\beq
\sum_{i=1}^r a_i {\pa \pre_k (a) \over \pa a_i}=(2-bk) \pre_k (a),
\label{eq:40}
\eeq
we obtain the one-instanton contribution
\beq
\pre_1(\bar a)={i \over \pi b} \sum_{\alpha \in \Delta^+}
(\alpha ,\bar a)(\alpha , \tilde a^{(1)}). 
\label{eq:41}
\eeq
Since (\ref{kthinst}) gives 
\beq
\tilde a^{(1)}={1\over 4} {\pa^2 \bar a \over \pa t_r^2}
={1\over 4} {\pa^2 \bar a \over \pa u_r^2},
\eeq
{}further manipulations lead to 
\beqa
\pre_1(\bar a) &=&{i \over 4\pi b}\sum_{\alpha \in \Delta^+}
(\alpha ,\bar a) {\pa^2 (\alpha , \bar a) \over \pa u_r^2} \CR
&=& {i \over 4\pi b}\sum_{\alpha \in \Delta^+}
\left[ \half {\pa^2 (\alpha , \bar a)^2 \over \pa u_r^2}
-\left( {\pa (\alpha , \bar a) \over \pa u_r} \right)^2 \right] \CR
&=& {1 \over 4\pi i b} \sum_{\alpha \in \Delta^+}
\left( {\pa (\alpha , \bar a) \over \pa u_r} \right)^2,
\label{eq:43}
\eeqa
where in the second line we have used (\ref{2ji}) and $\pa u_1/\pa u_r =0$.
Notice here that
\beq
{1\over h}\sum_{\alpha \in \Delta^+} (\alpha ,\alpha_i)(\alpha ,\alpha_j)=
K_{ij},
\eeq
then we find
\beq
\cF_{1}(\Ba)={1\over 8\pi i} \sum_{i,j=1}^r {\pa\Ba_{i}\over \pa u_{r}}K_{ij}
{\pa\Ba_{j} \over \pa u_{r}}.
\eeq

Let us introduce  the Jacobian determinant $D$, as well as  
its $(i,j)$-th cofactor $D_{ij}$,  
of the map from $(\Ba_{1}, \cdots, \Ba_{r})$ 
to $(u_{1}(\Ba), \cdots, u_{r}(\Ba))$ :
\beqa
D&=&{\rm det} \left( {\pa u_{m} \over \pa \Ba_{n}} \right), \CR
D_{i j}&=&(-1)^{i+j} {\rm det} {}_{m\neq i,n\neq j}
\left( {\pa u_{m} \over \pa \Ba_{n}} \right), 
\eeqa
in terms of which we have 
\beq
{\pa \Ba_{i} \over \pa u_{r}}={D_{r i}\over D}.
\eeq
Thus we finally arrive at
\beq
\cF_{1}(\Ba)={1 \over 8\pi i}\sum_{i,j=1}^r
 {D_{r i}K_{i j} D_{r j}\over D^{2}}.
\label{eq:pre1}
\eeq

We now compare our results with those derived by microscopic instanton
calculations. The general expression for the microscopic one-instanton
term $\pre_1^{micro}$ in A-D-E Yang-Mills theory is found in \cite{ItSa}.
The result reads
\beq
\pre_1^{micro}={1\over 2^{b/2}\pi i}F_1(a),
\label{eq:prem}
\eeq
where
\beq
F_1(a)=
{\sum_{\alpha \in \Delta^+} 
\prod_{\alpha^0 \in \Delta^+; (\alpha ,\alpha^0)=0}
(a,\alpha^0)^2 \prod_{\alpha^1 \in \Delta^1(\alpha)}
(a,\alpha^1)(a, \alpha^1-\alpha) \over
\prod_{\alpha \in \Delta^+} (a,\alpha)^2}.
\label{F1micro}
\eeq
Here $\Delta^1(\alpha)$ denotes a set of roots which become the highest weights
of $SU(2)$ doublets when we embed an $SU(2)$ into $G$ by taking a positive
root $\alpha$ as an $SU(2)$ direction. $\pre_1(a)$ and $\pre_1^{micro}(a)$
should be related through
\beq
\pre_1(a) \La^b = \pre_1^{micro}(a) \La_{micro}^b,
\eeq
where $\La_{micro}$ is the dynamical scale associated with 
the Pauli-Villas regularization scheme. Hence what we have to show is that
\beq
F_1(\Ba)= 2^h\pi i \pre_1(\Ba) {\La^b \over \La_{micro}^b}.
\label{eq:rel}
\eeq

Before calculating the explicit form of $\cF_{1}(\Ba)$
we examine its singularity structure.
To do this we first wish to discuss some properties of the 
Jacobian.
The Casimirs $(u_{1}(\Ba), \ldots, u_{r}(\Ba))$ are invariant 
under the Weyl reflection $\sigma_{\beta}$ with respect to any root 
$\beta$
\beq
u_{k}(\Ba)=u_{k}(\Bb),
\label{eq:48}
\eeq
where
\beq
\Bb=\sigma_{\beta}(\Ba)=\Ba-{2 (\beta, \Ba)\over (\beta,\beta)}\beta.
\eeq
{}From (\ref{eq:48}) we get
\beqa
{\pa u_{k}(\Ba)\over \pa \Ba_{i}}&=&
\sum_{j=1}^{r}{\pa\Bb_{j}\over \pa \Ba_{i}} {\pa u_{k}(\Bb)\over \pa \Bb_{j}}
\CR
&=&  {\pa u_{k}(\Bb)\over \pa \Bb_{i}}
-{2(\alpha_{i},\B)\over (\beta,\beta)}
\sum_{j=1}^{r}\beta_{j} {\pa u_{k}(\Bb)\over \pa \Bb_{j}},
\eeqa
where $\Bb_{i}$ is defined by $\Bb=\sum_{i=1}^{r}\Bb_{i}\alpha_{i}$.
If $\Ba$ is on the hyperplane defined by $(\beta,\Ba)=0$, we get $\Bb=\Ba$ 
and 
\beq
\sum_{j=1}^{r}\beta_{j} {\pa u_{k}(\Ba)\over \pa \Ba_{j}} =0.
\label{eq:51}
\eeq
This means that  the expression 
$\sum_{j=1}^{r}\beta_{j} {\pa u_{k}(\Ba)\over \pa \Ba_{j}}$,
which is a polynomial in $\Ba$'s,   
is divided by $(\beta,\Ba)$,  and hence takes  the form 
\beq
\sum_{j=1}^{r}\beta_{j} {\pa u_{k}(\Ba)\over \pa \Ba_{j}}=
(\beta,\Ba) f_{k,\beta}(\Ba)
\label{eq:52}
\eeq
with  $f_{k,\beta}(\Ba)$ being a certain polynomial in $\Ba$'s \cite{TeYa}. 
{}From (\ref{eq:52}) we find that the Jacobian determinant $D$
is also divided by $(\beta,\Ba)$ for any positive root $\beta$.
Since the sum of exponents $\sum_{i=1}^{r} e_{i}$ is equal to the number 
of positive roots, 
$D$ is shown to be expressed as \cite{Hum}
\beq
D=c \prod_{\alpha \in \Delta^{+}} (\alpha ,\Ba),
\eeq
where $c$ is a certain constant which depends on the choice of 
the normalization of the Casimirs.

It is now observed that the denominator of the RHS of (\ref{eq:pre1}) 
vanishes when $(\A,\Ba)=\ep\rightarrow0$ for a positive root 
$\A$. Due to invariance of $\cF_{1}(\Ba)$ under the Weyl transformation,
one may choose $\A$ as a simple root $\A_{1}$ without loss of generality.
Choosing $\beta=\A_{1}$ in (\ref{eq:52}) we find that 
\beq
{\pa u_{k}(\Ba)\over \pa \Ba_{1}}=\ep f_{k,\A_{1}}(\Ba)\sim O(\ep)
\hskip10mm k=1,\cdots ,r 
\eeq
and the other ${\pa u_{k}(\Ba)\over \pa \Ba_{i}}$ ($i\neq1$) are $O(1)$.
We have $D_{r1}\sim O(1)$ and $D_{ri}\sim O(\ep)$
($i\neq1$).
Thus, in the numerator of (\ref{eq:pre1}) only the term 
$D_{r1}K_{11}D_{r1}$ is of order one and contributes to the most singular 
term in $\cF_{1}(\Ba)$.
In the limit $\ep\rightarrow 0$, $D_{r1}$ behaves as
\beq
D_{r1}=d_{1} \prod_{\A\in \Delta'}(\A,\Ba)+O(\ep),
\eeq
where $d_{1}$ is a constant and $\Delta'$ is a subset of 
$\Delta^{+}$, which consists of the
linear combinations of simple roots $\A_{2}, \cdots , \A_{r}$.
We note that the positive root system can be decomposed into the sets
$\Delta_{i}$  whose element $\A$ satisfies $(\A_{1},\A)=i$:
\beq
\Delta^{+}=\Delta_{2}\cup \Delta_{1}\cup \Delta_{-1}\cup\Delta_{0}.
\eeq
It is easy to see that $\Delta_{2}=\{ \A_{1} \}$, 
$\Delta_{1}=\Delta^{1}(\A_{1})$,
$\Delta_{-1}=\{ \alpha-\alpha_{1}; \alpha\in\Delta_{1}\}$ and
$\Delta'=\Delta_{-1}\cup \Delta_{0}$.

To summarize, in the limit $\ep\rightarrow 0$, we obtain
\beq
\cF_{1}(\Ba)={1\over 8\pi i} 
{2 d_{1}^2 \over c^{2} \ep^{2}\prod_{\A\in\Delta^{1}(\A_{1})}(\A,\Ba)^{2}}
+\cdots .
\eeq
This is the same singularity structure which is expected from the microscopic
instanton result (\ref{eq:prem}). 
In view of  the holomorhic property of the prepotential, 
we thus conclude that the relation (\ref{eq:rel}) holds up to a
proportional constant. 
In order to determine this constant, we need to calculate 
the Jacobian determinant explicitly.
In the following we first express $D$ and $D_{i r}$
explicitly in terms of $\Ba_{i}$ for the $A_{r}$ and $D_{r}$ gauge groups
to work out (\ref{eq:rel}). We next proceed to the $E_6$ gauge group.

\subsection{$A_{r}$ gauge group}

For the group $A_{r}$, we consider the $(r+1)$-dimensional representation, 
which is realized by the traceless $(r+1)\times (r+1)$ matrices. 
It is therefore convenient to consider the $gl(r+1)$ case, where the
Cartan part $(\Ba_{1},\cdots, \Ba_{r+1})$ is generic, and then impose
the traceless condition $\sum_{i=1}^{r+1}\Ba_{i}=0$.
\footnote[4]{In subsections 4.1 and 4.2 our use of notations $\Ba_i$ is
slightly different from that in the other sections.}
As is shown in \cite{ItSa}, $F_1(\Ba )$ in (\ref{F1micro}) is then expressed as
\beq
F_{1}(\Ba)={\sum_{i=1}^{r+1}
\Delta(\Ba_{1}, \cdots,\widehat{\Ba_{i}},\cdots,\Ba_{r+1})
\over 
2 \Delta(\Ba_{1}, \cdots,\Ba_{r+1})},
\label{eq:ar}
\eeq
where $\widehat{\Ba_{i}}$ means to exclude $\Ba_{i}$, and
\beq
\Delta(\Ba_{1}, \cdots,\Ba_{n})=\prod_{1\leq i<j\leq n}(\Ba_{i}-\Ba_{j})^2.
\eeq

On the other hand, the Casimirs $u_{1},\cdots , u_{r+1}$ of $gl(r+1)$ are  
defined by the characteristic polynomial 
\beq
\prod_{i=1}^{r+1}(x-\Ba_{i})=x^{r+1}-\sum_{i=1}^{r+1}u_{i} x^{r+1-i}.
\eeq
These Casimirs are related to 
$s_{k}\equiv\sum_{i=1}^{r+1}\Ba_{i}^{k}$ via Newton's formula
\beq
s_{k}-u_{1}s_{k-1}-u_{2} s_{k-2}-\cdots -u_{k-1} s_{1}- k u_{k}=0,
\label{eq:newton}
\eeq
{}from which we get
\beq
{\pa u_{k}\over \pa \Ba_{i}}={1\over k} {\pa s_{k}\over \pa \Ba_{i}}
+\sum_{m\geq 2} \sum_{j_{1}+\cdots+j_{m}=k} d_{j_{1},\cdots, j_{m}}
s_{j_{1}}\cdots s_{j_{m-1}}{\pa s_{j_{m}}\over \pa \Ba_{i}},
\eeq
where $d_{j_{1},\cdots, j_{m}}$ are constants. Then we obtain
\beq
D={1\over (r+1)!}\det \left({\pa s_{i}\over \pa \Ba_{j}}\right) .
\eeq
Notice that this is nothing but the Vandermonde determinant 
and is shown to be
\beq
D=(-1)^{(r+1)r/2}\prod_{i<j}(\Ba_{i}-\Ba_{j}).
\eeq
In a similar way one obtains
\beq
D_{r+1 k}=(-1)^{r(r-1)/2}\prod_{{i<j}\atop{i,j\neq k}}(\Ba_{i}-\Ba_{j}).
\eeq
Hence we have
\beq
\cF_{1}(\Ba)={1 \over 8\pi i} {\sum_{i=1}^{r+1}
\Delta(\Ba_{1}, \cdots,\widehat{\Ba_{i}},\cdots,\Ba_{r+1})
\over 
\Delta(\Ba_{1}, \cdots,\Ba_{r+1})}.
\label{eq:arB}
\eeq
Putting (\ref{eq:ar}) and (\ref{eq:arB}) into (\ref{eq:rel}) yields
\beq
\La^b_{micro}=2^{r-1} \La^b.
\label{ratio}
\eeq
with $b=2(r+1)$. This agrees with the value evaluated in \cite{ItSa}.

\subsection{$D_{r}$ gauge group}

For the group $D_{r}$, let $\Ba_i$ be the skew eigenvalues of the matrix
in the $2r$-dimensional representation, then it is shown that 
$F_1(\Ba )$ in (\ref{F1micro}) becomes \cite{ItSa}
\beq
F_1(\Ba )={2 \sum_{i=1}^{r}\Ba_{i}^{2}
\Delta(\Ba_{1}^{2}, \cdots,\widehat{\Ba_{i}^{2}},\cdots,\Ba_{r}^{2})
\over 
\Delta(\Ba_{1}^{2}, \cdots,\Ba_{r}^{2})}.
\label{eq:73}
\eeq
To rewrite (\ref{eq:pre1}) we examine the characteristic polynomial for the 
$2r$-dimensional representation
\beq
\prod_{i=1}^{r}(x^2-\Ba_{i}^2)
=x^{2r}-\sum_{i=1}^{r} \tilde{u}_{i} x^{2r-2i},
\eeq
where $\tilde{u}_{i}$'s are related to the Casimirs $u_{1}, \cdots , u_{r}$ 
of $D_{r}$ by the formulas: $\tilde{u}_{i}=u_{i}$ ($i=1,\cdots, r-2$),
$\tilde{u}_{r-1}=u_{r}$ and $\tilde{u}_{r}=u_{r-1}^2$.
Using $u_{r-1}=\Ba_{1}\cdots \Ba_{r}$, we find that
\beq
\det \left( {\pa \tilde{u}_{i}\over \pa\Ba_{j}}\right)
=-2u_{r-1}D.
\label{eq:drvm}
\eeq
Since the LHS of (\ref{eq:drvm}) becomes 
$(-1)^{r(r-1)/2}\, 2^{r}u_{r-1} 
\prod_{1\leq i<j\leq r}(\Ba_{i}^{2}-\Ba_{j}^{2})$ we obtain
\beq
D=-(-1)^{r(r-1)/2}\, 2^{r-1}\prod_{1\leq i<j\leq r}(\Ba_{i}^{2}-\Ba_{j}^{2}).
\label{eq:dr1}
\eeq
Similarly 
\beq
D_{r k}=-(-1)^{(r-1)(r-2)/2}\, 2^{r-2}\, \Ba_{k}
\prod_{{1\leq i<j\leq r}\atop{i,j\neq k}}(\Ba_{i}^{2}-\Ba_{j}^{2}).
\label{eq:dr2}
\eeq
Thus $\cF_{1}(\Ba)$ is rewritten as
\beq
\cF_{1}(\Ba)={1 \over 8\pi i}  {\sum_{i=1}^{r}\Ba_{i}^{2}
\Delta(\Ba_{1}^{2}, \cdots,\widehat{\Ba_{i}^{2}},\cdots,\Ba_{r}^{2})
\over 
4 \Delta(\Ba_{1}^{2}, \cdots,\Ba_{r}^{2})}.
\label{eq:78}
\eeq
The ratio of the dynamical scales is determined from (\ref{eq:73}),
(\ref{eq:78}) and (\ref{eq:rel}). We get
\beq
\La^b_{micro}=2^{2r-8} \La^b
\label{eq:79}
\eeq
with $b=2(r-1)$, in agreement with \cite{ItSa}.

\subsection{$E_6$ gauge group}

For $E_{6}$, we may construct the spectral curve from the fundamental 
representation ${\bf 27}$ \cite{LW}.
The superpotential reads
\beq
W_{E_6}^{\bf 27}={1\over x^3} \left( q_1 \pm p_1\sqrt{p_2}\right)-u_6 ,
\label{e6curve}
\eeq
where
\beqa
&&q_1=270 x^{15}+342 u_{1} x^{13}+162 u_{1}^2 x^{11}-252 u_{2} x^{10}
+(26 u_{1}^3+18 u_{3}) x^{9}-162 u_{1} u_{2} x^{8} \CR
&& \quad \quad  +(6 u_{1} u_{3} -27 u_{4}) x^{7}
-(30 u_{1}^2 u_{2}-36 u_{5}) x^{6}
+(27 u_{2}^2 -9 u_{1} u_{4}) x^{5} \CR
&& \quad \quad -(3 u_{2} u_{3}-6 u_{1} u_{5}) x^{4}
-3 u_{1} u_{2}^2 x^3-3 u_{2} u_{5} x-u_{2}^3, \CR
&& p_1=78 x^{10}+60 u_{1} x^{8} +14 u_{1}^2 x^{6}-33 u_{2} x^{5}
+2 u_{3} x^{4}-5 u_{1} u_{2} x^{3}-u_{4} x^{2}-u_{5} x-u_{2}^2,  \CR
&& p_2=12 x^{10}+12 u_{1} x^{8}+4 u_{1}^2 x^{6}-12 u_{2} x^{5}+u_{3}x^{4}
-4 u_{1} u_{2} x^{3}-2 u_{4} x^{2}+4 u_{5} x+u_{2}^2. \CR
&& 
\eeqa
The Casimirs are expressed in terms of $\Ba_{i}$ as follows
\footnote{We thank M. Noguchi and S. Terashima for providing us with
some $E_{6}$ data.}
\beqa
u_{1}&=&-{1\over 12} \chi_{2}, 
\quad u_{2}=-{1\over 60} \chi_{5},
\quad u_{3}=-{1\over 6} \chi_{6}+{1\over 6\cdot 12^{2}}\chi_{2}^{3}, \CR
u_{4}&=&-{1\over 40} \chi_{8}+{1\over 180} \chi_{2}\chi_{6}
-{1\over 2\cdot 12^{4}} \chi_{2}^{4}, \quad
u_{5}= -{1\over 7\cdot 6^{2}}\chi_{9}+{1\over 20\cdot 6^{3}}\chi_{2}^{2}
\chi_{5}, \CR
u_{6}&=& -{1\over 60} \chi_{12}+{1\over 5\cdot 6^{3}}\chi_{6}^{2}
+{13\over 5\cdot 12^{3}}\chi_{2}\chi_{5}^{2} \CR
& & +{5\over 2\cdot 12^{3}}\chi_{2}^{2}\chi_{8}
-{1\over 3\cdot 6^{4}}\chi_{2}^{3}\chi_{6}
+{29\over 10\cdot 12^{6}}\chi_{2}^{6},
\label{e6casi}
\eeqa 
where $\chi_{n}={\rm Tr}\,\Phi^{n}$ and 
$\Phi={\rm diag}\left( 
(\lambda^{(1)},a),\cdots ,(\lambda^{(27)},a) \right)$.
The weight vectors $\lambda^{(i)}$ of ${\bf 27}$ are listed in \cite{SLA}.

The ratio $\LM^b/\LM_{micro}^b$ with $b=24$ for $E_6$ theory 
has not been known. Thus the
present analysis predicts the value of $\LM^b/\LM_{micro}^b$ from 
(\ref{eq:rel}). Using (\ref{e6casi}) we evaluate $\cF_{1}(\Ba)$ and 
$\cF_{1}^{micro}(\Ba)$ numerically in the limit $(\A_{1},\Ba)\rightarrow 0$, 
and find
\beq
{\LM^{24}\over \LM_{micro}^{24}}=2^{-8}\, 3^{6}.
\eeq
In a similar vein, our analysis now makes it possible to determine the
ratio $\LM^b/\LM_{micro}^b$ for $E_7$ and $E_8$ theory in principle. To do
this we need to write down the Casimirs $u_i$ in terms of $\bar a_i$.
However, this computation, though straightforward, requires more computer
powers than those available for us at present, and hence goes beyond the
scope of this paper.

\section{Logarithmic solutions of the Picard-Fuchs equations}

In this section we study the dual SW periods
\beq
a_{D i}(t)=\oint_{B_{i}}\lambda_{SW}
\eeq
defined by the $B$-cycles which have intersections 
$\# (A_{i}\cap B_{j})=\delta_{ij}$. In the weak-coupling region 
$a_{D i}$ may be evaluated  by using $a_{D i}=\pa \cF/\pa a_{i}$ and
(\ref{prepo}). One obtains
\beqa
a_{D i}(t)&=&\tau_{0}\sum_{j=1}^{r}K_{i j} a_{j}
+{i\over 2\pi}\sum_{\A\in\Delta^{+}}(\A,\A_{i}) (\A,a)
\left(\ln {(\A,a)^{2}\over \Lambda^{2}}+1\right)
+\sum_{k\geq1}{\pa \cF_{k}(a)\over \pa a_{i}}\Lambda^{bk} \CR
&=& \Ba_{D i}(t)+\sum_{k\geq1} \tilde{a}_{D i}^{(k)}(t) \Lambda^{bk},
\label{eq:ad1}
\eeqa
where we have
\beqa 
&& \Ba_{D i}(t) = \tau_{0}\sum_{j=1}^{r}K_{i j} \Ba_{j}
+{i\over 2\pi}\sum_{\A\in\Delta^{+}}(\A,\A_{i}) (\A,\Ba)
\left(\ln {(\A,\Ba)^{2}\over \Lambda^{2}}+1\right), \CR
&& \tilde{a}_{D i}^{(1)}(t) =  \tau_{0}\sum_{j=1}^{r}K_{i j}
 \tilde{a}^{(1)}_{j}
+{3i\over 2\pi}\sum_{\A\in\Delta^{+}}(\A,\A_{i}) (\A,\tilde{a}^{(1)}) \CR
& & \hskip17mm  
+{i\over 2\pi}\sum_{\A\in\Delta^{+}}(\A,\A_{i}) (\A,\tilde{a}^{(1)})
\ln {(\A,\Ba)^{2}\over \Lambda^{2}}+{\pa \cF_1(\Ba)\over \pa \Ba_{i}}, \\
&& \hskip2mm  \cdots\cdots ,  \nonumber
\eeqa
upon substituting the weak-coupling expansion (\ref{aweak}) of $a_{i}$.
It is easy to see that 
\beq
{\cal D}^{2} \Ba_{D i}=0,
\eeq
and hence the scaling equation of (\ref{PFflat}) yields the recursion 
relations among $\tilde{a}_{D i}^{(k)}$.
The relation at $O(\Lambda^{b})$ is
\beq
{\cal D}^{2}\tilde{a}_{D i}^{(1)}-h^{2}{\pa^{2} \Ba_{D i}\over \pa t_{r}^{2}}
=0
\eeq
which reduces to the equation
\beq
{\pa \cF_{1}(\Ba) \over \pa \Ba_{i}}
={i\over 4\pi h}\sum_{\A\in\Delta^{+}} (\alpha , \alpha_i)
\left\{ {\pa^{2} (\A,\Ba)\over \pa t_{r}^{2}}+{h\over (\A,\Ba)} 
\left( {\pa (\A,\Ba)\over \pa t_{r}}\right)^{2} \right\}.
\label{eq:89}
\eeq
This result is compatible with (\ref{eq:40}), (\ref{eq:41}) 
and (\ref{eq:43}). In view of (\ref{eq:pre1}), (\ref{eq:89}) provides us with
non-trivial identities for the Jacobian determinant and its first derivatives.

Next we consider the Gauss-Manin system of (\ref{PFflat}) and, in particular,
concentrate on checking if the logarithmic terms in $\Ba_{D i}$
\beq
\sum_{\A\in\Delta^{+}}(\A,\A_{i}) (\A, \Ba)
\ln { (\A,\Ba)^{2}\over \Lambda^{2}}, \hskip10mm  i=1,\cdots, r
\label{eq:loga}
\eeq
satisfy the Gauss-Manin system.
Substitute (\ref{eq:loga}) in the Gauss-Manin system to obtain
\beq
\sum_{\A\in\Delta^{+}}
{(\A,\A_{i})\over (\A,\Ba)}  {\pa (\A,\Ba)\over \pa t_{m}} 
 {\pa (\A,\Ba)\over \pa t_{n}} 
=\sum_{\ell =1}^{r} {C_{m n}}^{\ell}(t) \sum_{\A\in\Delta^{+}}
{(\A,\A_{i})\over (\A,\Ba)}  {\pa (\A,\Ba)\over \pa t_{\ell}} 
 {\pa (\A,\Ba)\over \pa t_{r}}, 
\label{eq:gmlog}
\eeq
where the fact that $\Ba_{i}$ obey (\ref{PFflat}) has been taken into
account. To proceed a little further we feel it more appealing to express 
(\ref{eq:gmlog}) in terms of the LG variables. For this, recall that, 
given a representation ${\cal R}$ of $G$, $\Ba_i$ and the roots $e^{(b)}$
of the characteristic polynomial (\ref{charapol}) are 
related by (\ref{eq:15}).
Choosing appropriate  weights $\lambda^{(b)}$ we express $\Ba$ in terms
of $e^{(b)}$ 
\beq
\Ba_{i}=\sum_{b} S_{i b}\, e^{(b)},
\eeq
where $S$ is an invertible $r\times r$ matrix, so that 
\beq
{\pa \Ba_{i}\over \pa t_{m}}=-\sum_{b} S_{i b} 
{\phi_{m}(e^{(b)})\over W'(e^{(b)})}
\eeq
with the use of (\ref{1st}). 
It is also convenient to define
\beq
M_{b c}^{(i)}=
\sum_{j,k=1}^{r}
\sum_{\A\in\Delta^{+}}
{(\A,\A_{i}) (\A,\A_{j}) (\A,\A_{k}) 
\over (\A,\Ba)}S_{j b} S_{k c}.
\eeq
Eq.(\ref{eq:gmlog}) is then rewritten as
\beq
\sum_{b,c} M_{bc}^{(i)} {\phi_{m}(e^{(b)}) \phi_{n}(e^{(c)})
\over W'(e^{(b)})   W'(e^{(c)}) }
=\sum_{\ell=1}^{r} {C_{m n}}^{\ell}(t) \sum_{b,c} 
M_{bc}^{(i)}
 {\phi_{\ell}(e^{(b)})
\over W'(e^{(b)})   W'(e^{(c)}) }, \quad
i=1,\cdots, r 
\label{eq:gmlog1}
\eeq
which look like operator product expansions for chiral primaries.
Taking ${\bf 3}$ of $SU(3)$ and ${\bf 10}$ of $SU(5)$ as test examples, 
we have confirmed that this somewhat complicated relation holds correctly. 
These examples are non-trivial enough to convince us that 
the A-D-E Gauss-Manin system has the logarithmic solutions (\ref{eq:loga}). 
The general proof of the statement is highly desirable.

\section{Discussions and conclusions}

We have studied the low-energy effective prepotential 
of $N=2$ supersymmetric Yang-Mills theory with A-D-E gauge groups.
In the weak-coupling region we have solved the Picard-Fuchs equations for 
the SW periods, and constructed the solutions in the form of a power series.
Commutativity between the differential operators
${\pa \over \pa t_{r}}$  and  ${\cal L}_{ij}$ of the Gauss-Manin system 
(\ref{eq:comm}) is important for this construction. Then we have 
computed one-instanton corrections to the prepotential and shown
that the present results agree with  the ones obtained from the 
microscopic approach. 

The SW curves (\ref{curve}) are identical with the spectral curves for the
periodic Toda lattice. We stress that, for the fundamental representations,
while the spectral curves for $A_r$ and $D_r$ are of hyper-elliptic type,
they are not for $E_{6}$, $E_{7}$ and $E_{8}$. Let us show that the
hyper-elliptic ansatz for the SW curves fails in producing the relation
(\ref{eq:rel}) for the $E$-type gauge groups.
Take the fundamental representation $\cR_f$ with dimension $d$ of
$G=ADE$ where $d=(r+1), 2r, 27, 56, 248$ 
for $A_{r}$, $D_{r}$, $E_{6}$, $E_{7}$, $E_{8}$, respectively. 
The hyper-elliptic curve proposed in \cite{hyperell} takes the form
\beq
y^{2}=P_G^{\cR_f}(x)^{2}-\LM^{2h}x^{2d-2 h},
\label{eq:hyper}
\eeq
where the characteristic polynomial $P_G^{\cR_f}(x)$ 
is given in (\ref{charapol}). These curves are inferred from embedding 
the representation $\cR_f$ into the curves
for $N=2$ supersymmetric QCD with the $SU(d)$ gauge group and massless 
$N_{f}=2d-2h$ fundamental flavors. 

Corresponding to the curves (\ref{eq:hyper}), therefore,
one may write down the one-instanton contribution to the prepotential
\beq
{\cF}_1(a)\LM^{2h}={\LM^{2h}\over 2\pi i} 
{\sum_{i=1}^{d} b_{i}^{2d-2h}\Delta(b_{1},\cdots, \widehat{b_{i}},
\cdots, b_{d})
\over \Delta (b_{1}, \cdots, b_{d})},
\label{eq:pre2}
\eeq
with $b_{i}=(\lm^{(i)}, a)$, which is evaluated from the prepotential 
of $N=2$ $SU(N_{c})$ QCD with $N_{f}$ flavors \cite{KlNoPh}.
It is shown that (\ref{eq:pre2}) agree with those 
obtained from microscopic calculations in the case of $AD$-type gauge groups.
For the $E$-type gauge groups, however, the expression (\ref{eq:pre2})
behaves quite differently from the microscopic results.
This can be easily seen by investigating the singularity
structure of $\cF_{1}$. 
For the hyper-elliptic curves (\ref{eq:hyper}), the pole structure
of $\cF_{1}$ is characterized by the discriminant of the polynomial 
$P_G^{\cR_f}(x)$
\beq
\Delta (b_{1}, \cdots, b_{d})=\prod_{i<j}(\lm^{(i)}-\lm^{(j)}, a)^2 
\label{eq:sing1}
\eeq
which depends explicitly on the weight vectors $\lm^{(i)}$ of $\cR_f$.
On the other hand, the spectral-curve analysis predicts that the 
singularity structure of  $\cF_{1}$ is characterized by
\beq
\prod_{\A\in\Delta^{+}}(\A , a)^2
\label{eq:sing2}
\eeq
as is discussed in section 4. It is clear that
(\ref{eq:sing1}) and (\ref{eq:sing2}) do not agree with each other for the
$E$-type gauge groups. Therefore only the spectral curves associated with 
the periodic Toda lattice describe the non-perturbative corrections to the 
low-energy effective actions of $N=2$ Yang-Mills theory with the
$E$-type gauge groups. This is
in accordance with the result obtained by the method of $N=1$ confining
phase superpotentials \cite{TeYa}. An analogous situation
has been observed for $N=2$ $G_2$ theory \cite{LPG},\cite{It}.

In this paper we have found an interesting interplay between the 
A-D-E singularity theory and the non-perturbative weak-coupling properties
of the SW-type solutions of $N=2$ Yang-Mills theory. It will be worth
studying the massless soliton points and  $N=2$ superconformal points 
\cite{scft} in the strong-coupling region of vacuum moduli space in a 
systematic manner based on the A-D-E singularity/LG point of view.


\vskip10mm
We are grateful to S. Hosono and Y. Yamada for stimulating discussions.
The work of S.K.Y. was partly supported by Grant-in-Aid for Scientific
Research on Priority Areas (Physics of CP Violation) from the Ministry
of Education, Science, and Culture of Japan.

\newpage


\end{document}